\documentclass{PoS}

\title{Searches for counterparts of gravitational waves at very high energies with H.E.S.S.}

\ShortTitle{Searches for counterparts of GW at VHE with H.E.S.S.}

\author{\speaker{Monica Seglar-Arroyo$^1$}\\
        \llap{$^1$}IRFU, CEA, Universit\'e Paris-Saclay, F-91191 Gif-sur-Yvette, France\\
        E-mail: \email{monica.seglar-arroyo@cea.fr}}

\author{Halim Ashkar$^1$, Simon Bonnefoy$^2$, Francois Brun$^1$, Kathrin Egberts$^3$, Matthias F\"u{\ss}ling$^2$, Clemens Hoischen$^3$, Thomas Murach$^2$, Stefan Ohm$^2$, Gerd P\"uhlhofer$^4$, Gavin Rowell$^5$, Fabian Sch\"ussler$^1$, Andrew Taylor$^2$ on behalf of the H.E.S.S. Collaboration\footnote{for collaboration list see PoS(ICRC2019)1177}\\
\llap{$^2$}DESY, D-15738 Zeuthen, Germany\\
\llap{$^3$}Institut f\"ur Physik und Astronomie, Universit\"at Potsdam, Potsdam, Germany\\
 \llap{$^4$}Institut f\"ur Astronomie und Astrophysik, Universit\"at T\"ubingen, T\"ubingen, Germany\\
 \llap{$^5$} School of Physical Sciences, University of Adelaide, Adelaide
5005, Australia}

\abstract{The recent milestones in multi-messenger astronomy have opened new ways to study the Universe. The unprecedented gravitational wave (GW) follow-up campaigns established the power that the combination of different messengers has to identify and study the nature and evolution of astrophysical phenomena. Here we focus on the search for high-energy gamma ray emission as electromagnetic counterpart of compact binary coalescences with the H.E.S.S. Imaging Air Cherenkov Telescopes (IACTs).

In this contribution, the optimized strategies developed specifically for the prompt follow-up of gravitational wave events with H.E.S.S are presented. As illustration, the successful observation campaigns up to this time will be described, including the ones during Observation Run O2 on the binary black hole (BH-BH) merger GW170814 and the binary neutron star (NS-NS) merger GW170817, and an update on recent events occurring during O3. Results of these searches are presented and the constraints that prompt observations can put on very-high-energy, non-thermal emission, are briefly discussed.  Finally, an outlook on further improvements for the gravitational waves follow-up program with H.E.S.S. will be provided.}

\FullConference{36th International Cosmic Ray Conference -ICRC2019-\\
		July 24th - August 1st, 2019\\
		Madison, WI, U.S.A.}

\begin{document}


\section{Introduction}
In the last years, the multi-messenger astronomy accomplishments have shown how powerful this approach is to study astrophysical phenomena in the Universe. Unprecedented observation campaigns in the entire electromagnetic spectrum and in several messengers were launched after the first detection of a gravitational wave (GW) produced by a neutron star merger. This event showed that multi-messenger observations enables the characterization of the GW source by first finding its localization and then, by putting constraints on the processes that the source undergoes. 

Nevertheless, the search of an electromagnetic counterpart in gravitational wave triggers is non trivial, as the localization of compact binary coalescences detected by interferometers on Earth can be as coarse as hundred or thousands of square degrees \cite{prospects}. An additional challenge in GW follow-up is related to the latency of the follow-up, as in order to provide a prompt reaction to the alert, an efficient strategy should be in place. Furthermore, the detection technique in IACTs limits the duty cycle of the instrument, due to the good atmospheric conditions and darkness required during observations. Efforts to increase the H.E.S.S. duty cycle are ongoing, where observations under faint moon light conditions have been considered.

Dedicated strategies to follow-up gravitational wave events have been set up by H.E.S.S. in order to overcome the mentioned challenges. These enabled the follow-up of events during O2, where the H.E.S.S. experiment performed the first ground-based observation of the lately identified counterpart of the NS-NS merger, GW170817, and the follow-ups of the BH-BH merger GW170814, as well as during O3, on the BH-BH merger S190512at. 

In this work, the past, present and future of the gravitational-wave follow-up strategy of the H.E.S.S. experiment is described and presented in Section \ref{s0}. In Section \ref{s1} the algorithms used for the optimization of the prompt response to GW alerts are described, as well as the implementation in the H.E.S.S. alert system. Later, in Section \ref{s2}, the GW follow-up observations performed during O2 are presented and results are discussed. The GW-observations performed so far during O3 are presented in Section \ref{s3}. To sum up, in Section \ref{s4} an outlook on the gravitational wave program with H.E.S.S. will be given.

\section{The H.E.S.S. experiment}
\label{s0}

The High Energy Stereoscopic System (H.E.S.S.) experiment is a stereoscopic system consisting of five IACTs located at 1800 meters a.s.l. in the Khomas region, Namibia. The H.E.S.S. I array started operations in 2004, when the telescope array consisted in four 12m telescopes (CT1-4) arranged in a square of 120 m side length, whose field-of-view (FoV) is of $\sim$5 deg. An upgrade to H.E.S.S. II followed in 2012, when a fifth, 28m-telescope (CT5), with a FoV of $\sim$3.2 deg, was added to the center of the array. In its final phase, H.E.S.S. is sensitive to gamma-ray energies from $\sim$ 30 GeVs to about 100 TeV. 

The detection of very-high energy photons with the H.E.S.S. instrument is performed by collecting the Cherenkov radiation cascade emitted by the charged particles that are created when a primary photon interacts with particles present in the Earth's atmosphere. The main background source of this detection technique are air showers initiated by charged cosmic rays, which are much more numerous than photon-induced air showers.  Background discrimination is performed first at the level of the data acquisition using a two-level trigger, \textit{local} at each camera and \textit{central} between the triggered telescopes, and further discrimination is performed at the level of the reconstruction.  Details of the reconstruction of the showers and the discrimination technique can be found for the \textit{Model} analysis used in the presented work in \cite{model} and for the independent analysis chain used for cross-checks, \textit{ImPACT}, in \cite{impact}.

The configuration of the analysis of the H.E.S.S. II data varies depending on the number of telescopes involved in the data acquisition and the physical target. Throughout the analyses presented here, the \textit{monoscopic} mode has been chosen, making use of the data obtained by the  28 m telescope, in order to concentrate on lower energies.


\section{Optimized prompt follow-up of gravitational wave events with H.E.S.S.}
\label{s1}
\subsection{General observation scheduling considerations in IACTs}
The first step in order to schedule GW follow-up observations is account for the general constraints on observations of the H.E.S.S. telescopes, which apply to any observed source.
\begin{itemize}
\item \textbf{Visibility constraints:} The detection technique in IACTs, where the atmosphere acts as a part of the calorimeter of the instrument, requires clear skies and minimal light condition, including neither sun light nor moon light, and reduced values of light pollution during data acquisition. Nevertheless, the observation conditions can be slightly loosened as it is the case for Moonlight Observations. Although these type of observations present higher energy threshold and are less sensitive than \textit{standard} observation, these are indeed moderate effects compared to the science motivated discovery potential (Figure \ref{MoonLightObs}). 

\begin{figure}
\centering
 \includegraphics[width=12 cm]{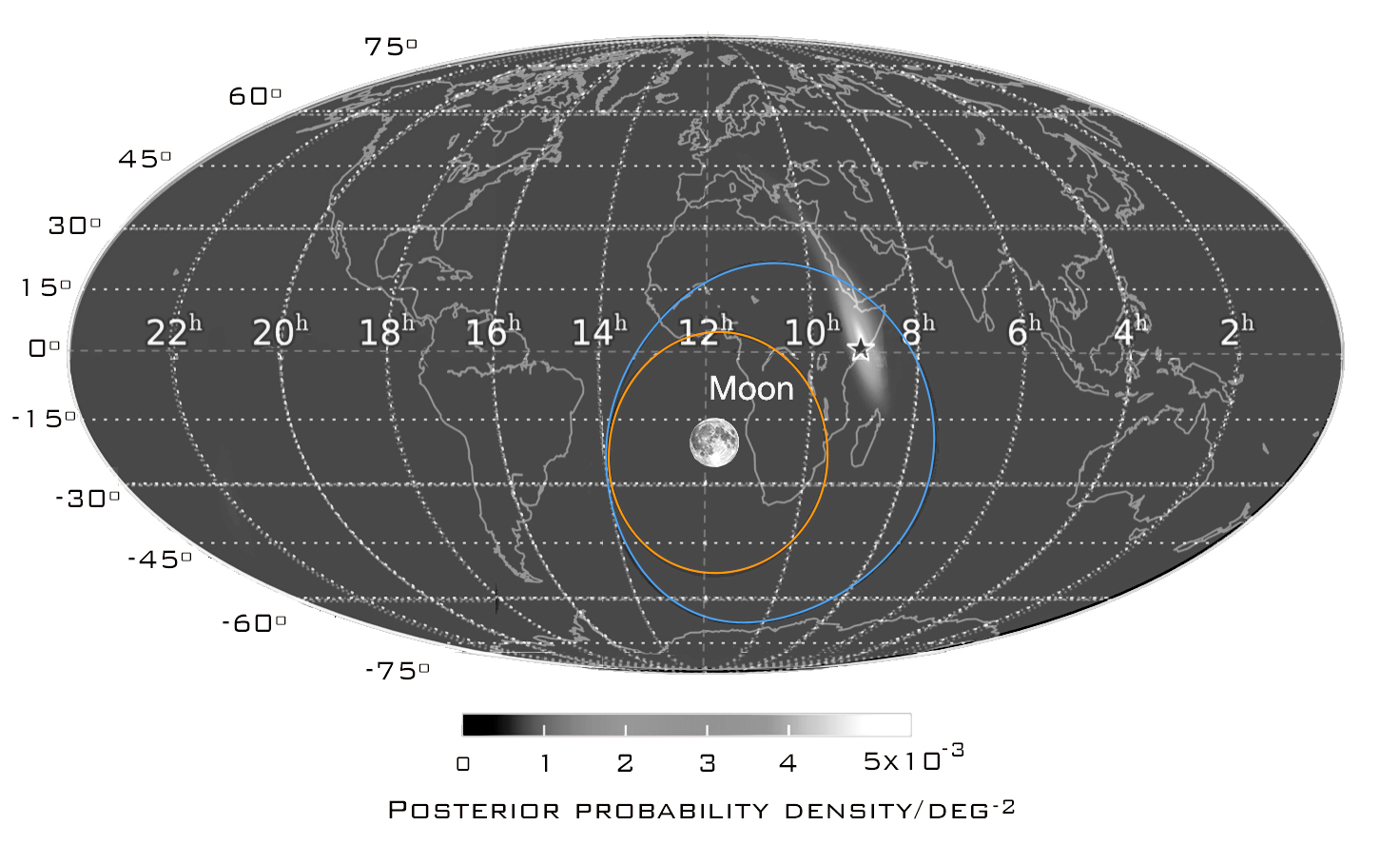}
\caption{Illustration of the Moon, the H.E.S.S. FoV at $\theta_{zen}=$60 in blue and a gravitational wave simulation from the GWCOSMoS database \cite{GWCOSMOS}, where orange corresponds illustrates a the minimum angular distance that fulfills the angular separation condition. We observe that in this case GW-follow up observations could be scheduled.}\label{MoonLightObs}
\end{figure}

\item \textbf{Zenith angle optimisation:} In IACTs, the energy threshold of an observation depends on the zenith angle under which the source is observed. This is due to the absorption of Cherenkov shower light during its passage through the atmosphere, which is larger for larger zenith angles. 
Since the high-energy counterpart of NS-NS mergers is expected to be a GRB \cite{blinn}, which show a soft spectrum as observed by \textit{Fermi}-LAT \cite{FermiCat2}, we prioritize the low energy domain in our observation scheduler. For this reason, the scheduling algorithms favor observations at low zenith angles.
\end{itemize}
\subsection{Probability selection algorithms}
A sequential order of the observations based on the covered probabilities in the GW localization maps, from the highest to the lowest, enhances the likelihood of covering the EM counterpart in a shorter period of time. Several techniques have been developed to guide the follow-up of gravitational wave events in a smart, efficient way, covering the most probable regions as fast as possible \cite{abadie2012}. The scheduling algorithms determine the pointing pattern of the telescopes by covering the regions where the source is most probably located at. This \textit{probability} can be defined in two different ways, corresponding to two scheduling algorithms, which are selected based on the characteristics of the GW event:
\begin{itemize}
    \item \textbf{2D-Scheduling algorithm}: The most straightforward approach to the scheduling problem is to use the two-dimensional localization posterior probability provided with the GW alerts. 
    \item \textbf{3D-Scheduling algorithm}: The horizon of the current gravitational wave interferometers is limited to the local Universe, so one of the possible approach is to search for known astrophysical objects which can potentially host the BNS merger. The reconstructed uncertainty volumes of the gravitational wave provided by the LIGO-Virgo collaboration, include estimates on the luminosity distance which can be combined with the position and redshift of known galaxies \cite{goingthedistance}.
\end{itemize}

\subsection{Implementation in the H.E.S.S. alert system}

\begin{figure}
\centering
 \includegraphics[width=14.5 cm]{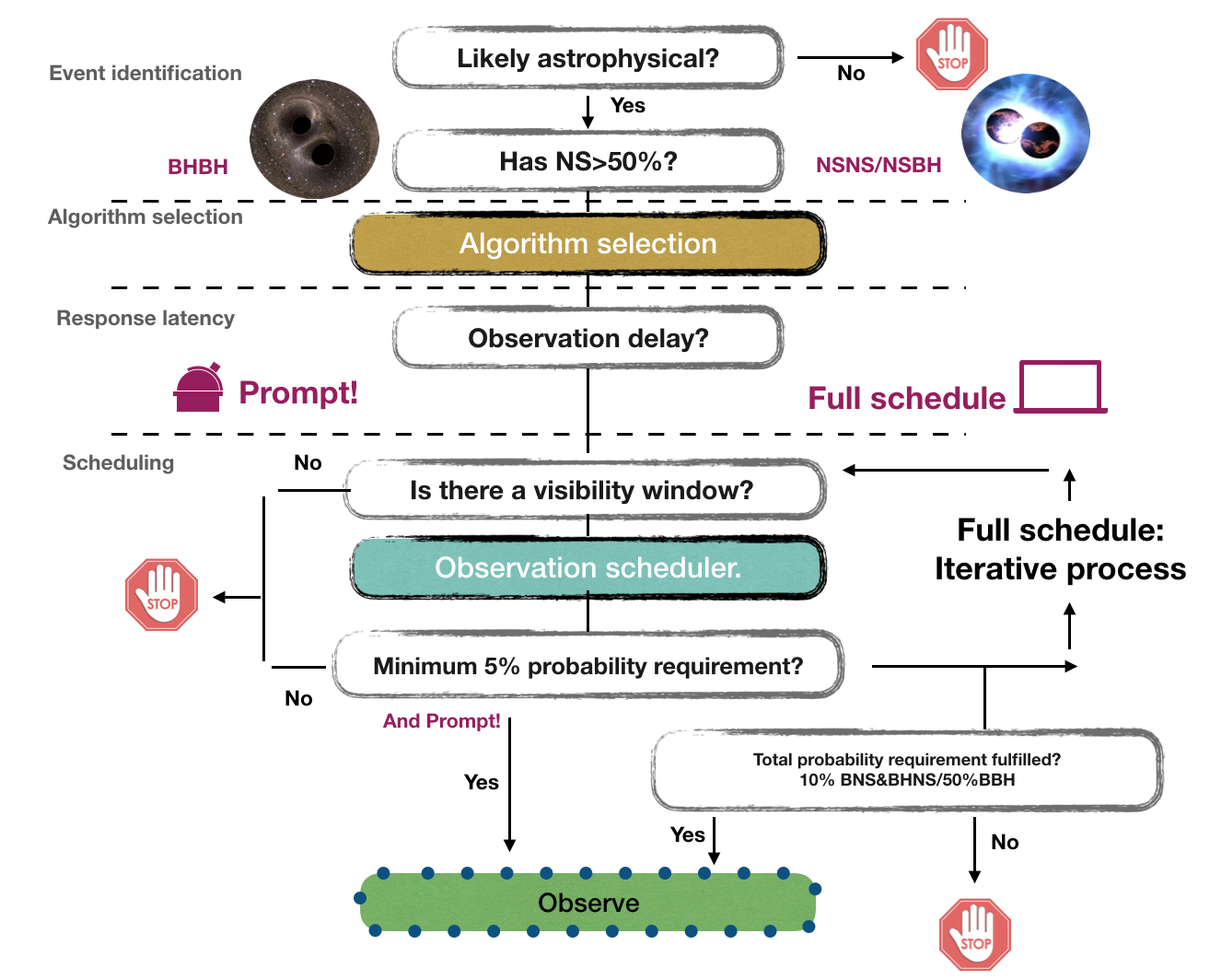}
\caption{Illustration of the response workflow of the VoAlerter to a gravitational wave alert.}\label{VoAlerterGW}
\end{figure}

The GW follow-up observation scheduler has been integrated in the H.E.S.S. online alert system, VoAlerter in the following. A schematic workflow of the response of the VoAlerter, which is based on the decision tree used to schedule observation offline, is presented in Figure \ref{VoAlerterGW}. The different decision levels can be grouped in event identification, algorithm selection, response latency, which depends the availability of observation window at the time of the alert reception, and scheduling of observations.

This approach was proven to be successful for the H.E.S.S. experiment to overcome the challenges that gravitational wave observation scheduling poses, as will be shown in the following sections.

\section{GW follow-ups with H.E.S.S. during O2}
\label{s2}
The H.E.S.S. experiment participated in multi-messenger follow-up of events during the O2 observation run, which lasted from November 2016 until August 2017. Two follow-up observation of gravitational wave triggers were performed on GW170814 and GW170817.

\subsection{The binary black hole merger GW170814}

\begin{figure}
\centering
 \includegraphics[width=7.68cm]{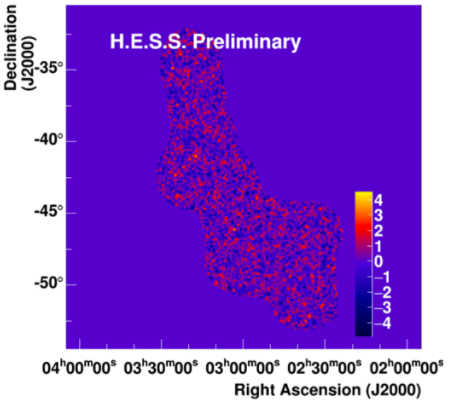}
 \includegraphics[width=7.33 cm]{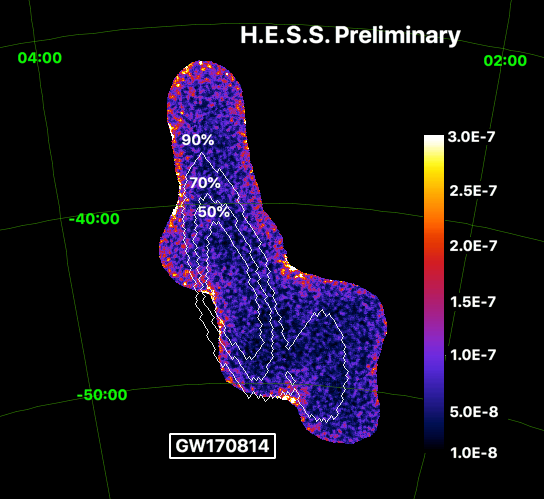}
\caption{(Left) Significance map of the gamma-ray emission obtained for the three nights of observations by H.E.S.S. of GW170814. Figure extracted from \cite{GW170814HESS}. (Right) Integral upper limits in units of ph m$^{-2}$ s$^{-1}$ between energies 250 GeV < E < 10 TeV. The uncertainty region of the reconstruction with LALInference \cite{veitch} of the sky localisation of the GW170814 is overlaid in white.}\label{GW170814}
\end{figure}

The H.E.S.S. telescopes followed up the first 3-interferometer detection of a gravitational wave, produced by a binary black hole merger, which was detected on August 14 2017, GW170814 \cite{GW170814}. The 90\% credible region for the event localization of this BH-BH expands over 60 deg$^2$ in the southern sky. Due to visibility constraints, observations got delayed and were scheduled throughout three nights, starting in August 17 at 00:10 UTC. A total of eleven 28-minutes observations were obtained with CT5 and three of the CT1-4 small telescopes. Each of the observations reached a sensitivity of about 20\% of the flux from the Crab nebula at 5 sigma. The H.E.S.S. observations cover around 90\% of the localisation of the final reconstruction.
 The significance sky map which results from a \textit{Model} analysis is shown in Figure \ref{GW170814}, where no significant gamma-ray emission is observed \cite{GW170814HESS}. Integral upper limits constraining gamma-ray emission from the remnant of the binary coalescence have been derived assuming a power-law spectrum of E$^{-2}$ for energies between 250 GeV < E < 10 TeV.

\subsection{The binary neutron star merger GW170817}

The first gravitational wave ever detected coming from the coalescence of a neutron star binary was recorded on August 17, 2017, at 12:41:04 UTC by the Advanced LIGO and Advanced Virgo interferometers \cite{GW170817Ligo}. An update of the BAYESTAR localization reconstruction \cite{sing} of the gravitational wave was provided at 17:54:51 UTC by the LIGO-Virgo collaboration and included cleaned data from LIGO-Livingston and Virgo data. The result of this joint analysis reduced the 90\% localization uncertainty of the GW event to $\sim$31 deg$^2$ and constraint the distance to 40$^{+8}_{-14}$ Mpc. The rapid reaction of the H.E.S.S. experiment allowed to schedule the first observations at 17:59 UTC, five minutes after the reception of the new localization of the neutron star merger, and continued until 19:30 UTC, time at which the last part of the localization region moved outside the H.E.S.S. FoV. A total of three observation runs were obtained. Then, after the remnant was pinpointed by optical instruments, the H.E.S.S observation schedule was modified accordingly for the following four nights, until the region was not observable any more by H.E.S.S. More information on the H.E.S.S. results of this follow-up observations can be found in \cite{HESSGW170817}.
\begin{figure}
\centering
 \includegraphics[width=7.5 cm]{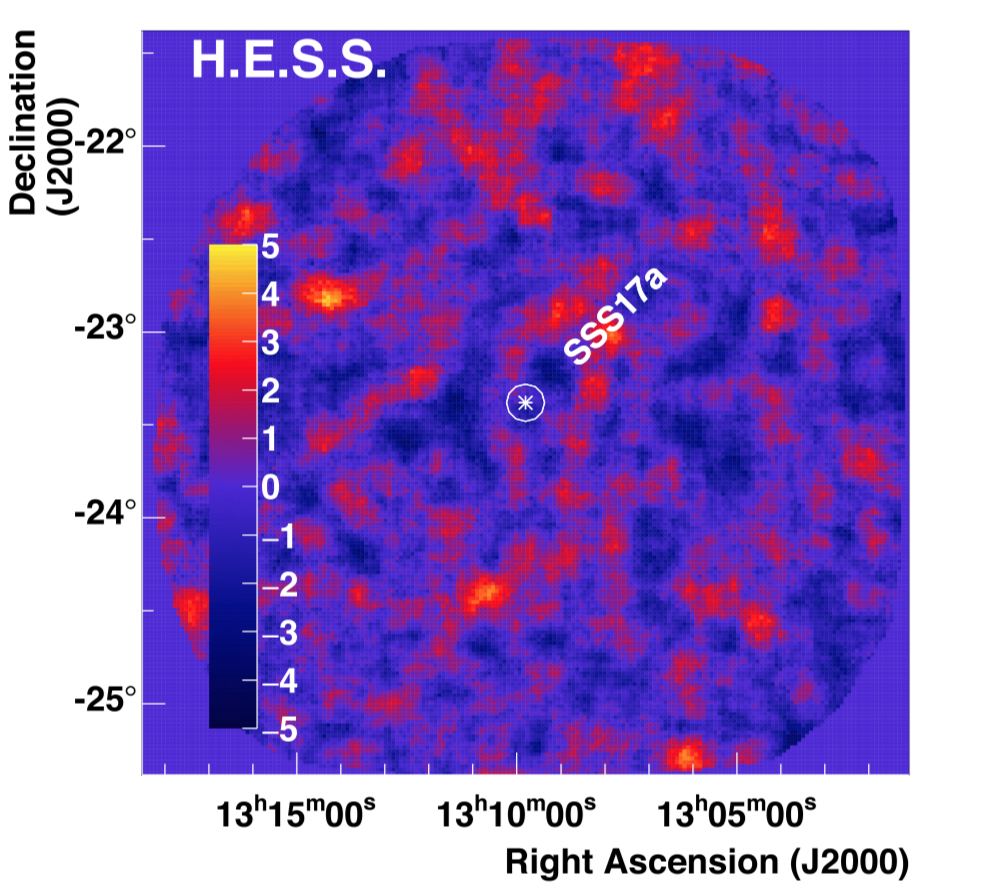}
 \includegraphics[width=6.1 cm]{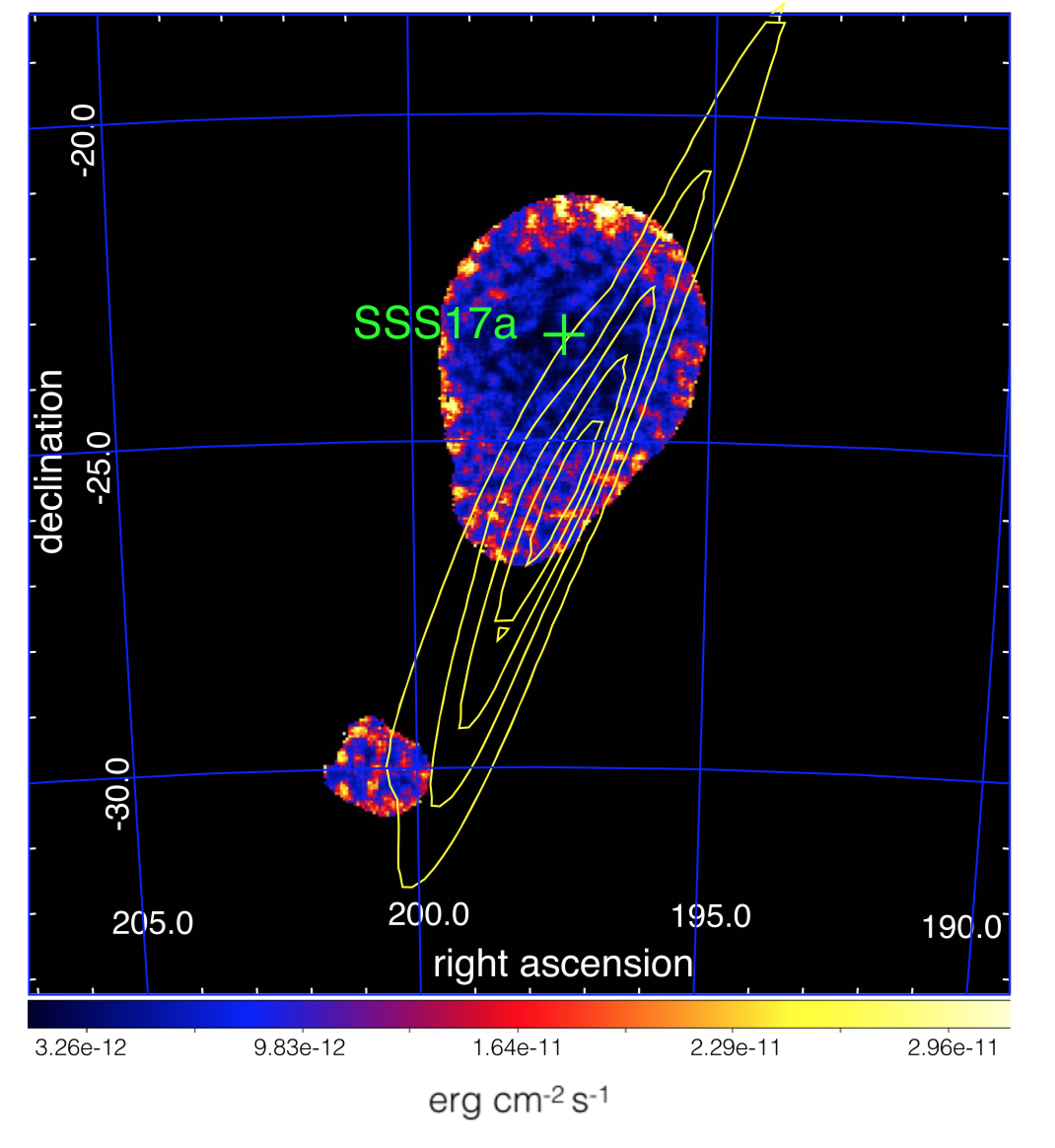}
\caption{(Left) Significance map of the gamma-ray emission for GW170817 combining all observations obtained during the H.E.S.S. monitoring campaign (Right) Sky map of the integral upper limits in the 270 GeV to 8.55 TeV energy range, obtained by assuming an E$^{-2}$ energy spectrum, derived from the H.E.S.S. follow-up observations of GW170817. The yellow contours correspond to the localization of the GW event reconstructed by the LALInference algorithm, distributed by the LIGO-Virgo Collaboration. Figures extracted from \cite{HESSGW170817}}\label{GW170817}
\end{figure}
In addition, the H.E.S.S. experiment performed a further monitoring of the GW170817 remnant at later times, between 120-250 days after merger, motivated by the disentanglement between  magnetic field of the ejecta and the maximum accelerated particle energy that gamma-ray observations can provide \cite{rodrigues}\cite{stefan}. 
\section{GW follow-ups with H.E.S.S. during O3}
\label{s3}

The H.E.S.S. experiment has attentively considered all the sent alerts since the beginning of O3. The characteristics of each gravitational wave have been considered in preparation of a potential decision starting H.E.S.S. follow-up observations.

\begin{figure}
\centering
 \includegraphics[width=7.1 cm]{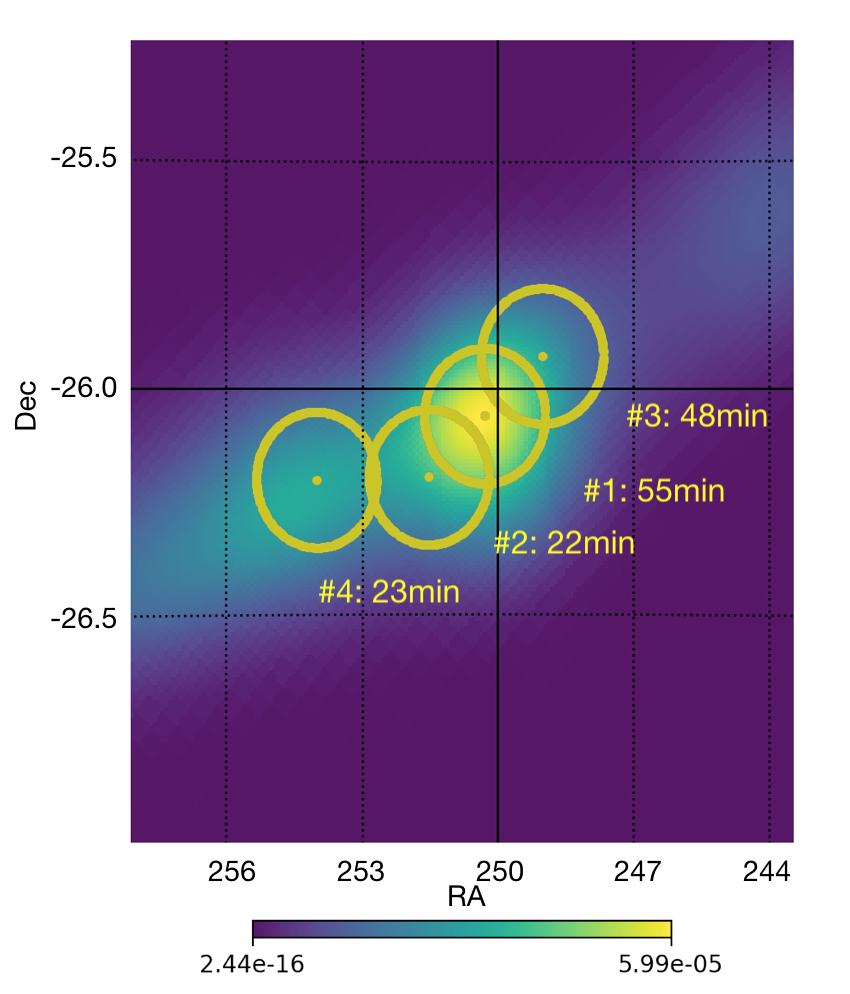}
\caption{H.E.S.S. follow-up scheduled observations of the BBH merger S190512at reconstructed by the BAYESTAR algorithm}\label{S190512}
\end{figure}

The first gravitational wave event followed-up by the H.E.S.S. collaboration during O3 is a coalescence of binary black holes at a distance of 1331 $\pm$ 341 Mpc. The BH-BH S190512at, which happened on May 12, 2019, was detected by the three interferometers of the GW network \cite{S190512at}. It presents a false alarm rate (FAR) of 1/16.6 years$^{-1}$, and it was relatively well localized, as the 90\% containment region expanded over 399 deg$^2$, in the prompt reconstruction. The H.E.S.S. experiment triggered follow-up observations on this event, with the motivation to perform a technical trial to check the follow-up chain.  Four observations were obtained, which are shown in Figure \ref{S190512}.

\section{Outlook}
\label{s4}

The gravitational wave follow-up strategy implemented by the H.E.S.S. was proven successful during O2, where the two follow-ups presented in this work were performed. The H.E.S.S. experiment continues to carefully follow GW alerts during O3 observation run, in order to react to potentially interesting GW events and put constraints on their very-high energy emission.

\section{Acknowledgements}

https://www.mpi-hd.mpg.de/hfm/HESS/pages/publications/auxiliary/HESS-Acknowledgements-2019.html

\end{document}